\documentclass[prd,aps,nofootinbib]{revtex4}

\usepackage{graphicx}% Include figure files
\usepackage{dcolumn}% Align table columns on decimal point
\usepackage{bm}% bold math

\def\x{{\bf x}}

\def\y{{\bf y}}

\begin{document}
\voffset = 0.3 true in
\topmargin = -1 true in % for Mac tetex

\title{Diagnostic Decays of the $X(3872)$}

\author{Eric S. Swanson}
\affiliation{
Department of Physics and Astronomy, University of Pittsburgh,
Pittsburgh PA 15260}

\vskip .5 true cm
\begin{abstract}
The unusual properties of the $X(3872)$ have led to speculation that it is
a weakly bound state of mesons, chiefly $D^0\bar D^{0*}$. Tests of
this hypothesis are investigated and it is proposed that
 measuring the $3\pi J/\psi$, $\gamma J/\psi$, $\gamma \psi'$, $\bar K K^*$, and $\pi\rho$
decay modes of the $X$  will serve as a definitive diagnostic of the molecule hypothesis.
\end{abstract}

%\pacs{}
\maketitle

\section{Introduction}

The discovery of a new charmonium state, the $X(3872)$, in $B$ decays\cite{belle,D0,CDF,babar1}\footnote{The
reported masses are $3872\pm 0.6 \pm 0.5$ (Belle); $3871.3 \pm 0.7 \pm 0.4$ (CDF); 
$3871.8 \pm 3.1 \pm 3.0$ (D0); and $3873.4 \pm 1.4$ (BaBar). The world average is $3871.9 \pm 0.5$ MeV.}
has caused some interest because the $X$ mass and its narrow width ($\Gamma < 2.3$ MeV, 90\% C.L.) do not agree well with quark model expectations.
The most likely candidates are 1D or 2P charmonium states; however, as Table I illustrates, the
1D states tend to lie below 3872 while the 2P states are somewhat above. The masses quoted in this
table are given by a simple nonrelativistic quark model with perturbative spin orbit and tensor
interactions and a smeared hyperfine interaction\cite{bgs}. The `model range' column of
this table is an average of several theoretical predictions\cite{BG}.

\begin{table}[h]
\caption{Quark Model $c\bar c$ Masses.}
\begin{tabular}{cccc}
state & expt (MeV) & mass (MeV) & model range (MeV) \\
\hline
$1^3D_3$ &   & 3806 & 3841(22) \\
$1^3D_2$ &   & 3800 & 3827(24) \\
$1^3D_1$ & 3770 & 3785 & 3803(25) \\
$1^1D_2$ &   & 3799  & 3821(32) \\
\hline
$2^3P_2$ &  &  3972  & 3990(25) \\
$2^3P_1$ &  &  3925  & 3957(28) \\
$2^3P_0$ &  &  3852  & 3903(40) \\
$2^1P_1$ &  &  3934  & 3964(20) \\
\hline
\end{tabular}
\end{table}

This situation has led to speculation that the $X$ could be a novel charmonium state,
such as a hybrid\cite{hybrids}, or a molecule\cite{moles}.  The former scenario is viewed 
as unlikely because lattice computations of the lightest charmonium hybrid yield masses
around 4400 MeV\cite{lgt}. 
The molecular hypothesis is bolstered by the proximity of the $X$ to the $D^0\bar D^{0*}$ 
threshold at 3871.2(7) MeV. Furthermore, the charged $D\bar D^*$, $\rho J/\psi$, and
$\omega J/\psi$ channels all lie within 7 MeV of the $X$ mass. Thus it is natural
to expect a strong admixture of these states in the $X$ (assuming it is indeed a resonance
and not a threshold enhancement). This scenario has been examined in Ref.\cite{ess} where
a 6 coupled channel model with quark exchange and pion exchange interactions was
diagonalised to determine the structure of the $X$.  The conclusion was that the only viable
molecular candidate is $J^{PC} = 1^{++}$ and is dominantly $D^0\bar D^{0*}$ with 
important but small admixtures of 
$\omega J/\psi$ and $\rho J/\psi$. This state was dubbed the $\hat \chi_{c1}$.
In this scenario, the decay $X \to \pi^+\pi^- J/\psi$ is permitted by the isospin symmetry
violating $\rho J/\psi$ component\footnote{Isospin violation is driven by
the 8 MeV difference between the  neutral and charged
$D\bar D^*$ channels. The resulting prediction of production of $\pi\pi$ via the $\rho$ is supported by the observed $\pi\pi$ spectrum\cite{belle,babar1}.} 
of the $X$.
 However, the $\omega J/\psi$
component can decay via $3\pi J/\psi$ and detecting the $X$ in this mode will be an important
test of this scenario.
We remark that isospin violating decay modes are a generic feature of molecular states
with binding energies comparable to the mass splittings of its constituent neutral and charged mesons.

Recent experimental effort has significantly narrowed the range of options available for the
structure of the $X$. The Belle collaboration\cite{choi} has measured the ratio of partial widths

\begin{equation}
{\Gamma(X \to \gamma \chi_{c1})\over \Gamma(X \to \pi^+\pi^- J/\psi)} < 0.89,\ {\rm 90\%\  C.L.}
\end{equation}
and stated that this contradicts expectations for the $\psi_2$ identification of the $X$\footnote{This assumes that $\Gamma(\psi_2 \to \pi^+ \pi^- J/\psi) \approx \Gamma(\psi(3770) \to \pi^+ \pi^- J/\psi) \approx 100$ keV.}. Similarly, there is a weak signal in $\gamma\chi_{c2}$, indicating that the $\psi_3$ identification is unlikely. Belle also reports that the helicity angle distribution of the $J/\psi$ in the $X$ final state rules
out $J^{PC} = 1^{+-}$. BaBar have not seen the $X$ in the $\eta J/\psi$
mode\cite{babar1} which is consistent with the molecular interpretation of the $X$. BES reports that 
the $X$ is not seen in initial state radiation production\cite{YMW}, implying that it is inconsistent 
with $J^{PC} = 1^{--}$.  Finally, CLEO has not detected the $X$ in untagged two 
photon production\cite{cleo}, which indicates that $J^{PC} = 0^{\pm +}$ and $2^{\pm +}$ are disallowed. 

Together, these observations imply that the only viable nonexotic quantum numbers below 
$J=3$ are $J^{PC} = 1^{++}$.  The remainder of this paper therefore concentrates on the
$2^3P_1$ $c\bar c = \chi'_{c1}$ and molecular $\hat \chi_{c1}$ assignments for the $X$.
Distinguishing these assignments is a task for experiment coupled with reliable theoretical
expectations.
The importance of measuring the $3\pi J/\psi$ mode arising from the
short range $\omega J/\psi$ component of the $\hat \chi_{c1}$ has already been mentioned.
Here we examine the utility of a novel annihilation decay and radiative decays of the $X$ as diagnostic tools. 

Radiative decays of the molecular $\hat \chi_{c1}$ to charmonium may occur via vector
meson dominance in the $\rho J/\psi$ or $\omega J/\psi$ components of the $X$. Thus
$\gamma J/\psi$ is the only possible final state available to this mechanism. An 
additional mechanism has the light quarks coupling to the final state photon from 
the neutral and charged $D\bar D^*$ components of the $\hat \chi_{c1}$. This diagram
permits coupling to a variety of charmonia in the final state. Specific computations,
presented in the next section, reveal that measurements of the rates $X \to \gamma J/\psi$ and $X \to \gamma \psi''$ will definitively distinguish the charmonium from the molecular
options for the $X(3872)$. 

The annihilation process $X \to K\bar K^*$ is studied in Section 3. This decay of the
molecular state may occur via single gluon exchange but is strongly suppressed by
the weakly bound nature of the $\hat \chi_{c1}$. 
The multigluon exchange process which 
contributes to $c\bar c \to K\bar K^*$  is also suppressed but comparisons with typical
charmonia indicate that charmonium rates for this reaction should be three orders of
magnitude larger than that for the molecular state. A novel feature of the molecular 
decay is that production of charged kaons is driven by the dominant neutral $D\bar D^*$ component
of the $\hat \chi$ and therefore should be larger than decay into neutral kaons.

\section{Radiative Decays of the $X$}

The primary mechanisms  for radiative decays of the $\hat\chi_{c1}$ are
via vector meson dominance  (Fig. 1) and light quark annihilation (Fig. 2). The vector
meson dominance (VMD) process proceeds via the $\rho J/\psi$ and $\omega J/\psi$ components
of the $\hat\chi_{c1}$ and thus contributes to $\Gamma(X \to \gamma J/\psi)$. It is clear that the amplitude must be proportional to the light vector meson wavefunction at the origin
and to the $\hat\chi$ wavefunction for the channel in question evaluated at the recoil momentum. The specific result is

\begin{equation}
\Gamma_{\rm VMD} = {4\over 27} \alpha {q E_\psi \over m_\chi} |\psi_\omega(r=0)|^2  \left( Z_{\omega\psi}\phi_{\omega\psi}(q) + 3 Z_{\rho\psi} \phi_{\rho\psi}(q)\right)^2.
\label{vmd}
\end{equation}
Here it is  assumed that the $\rho$ and $\omega$ wavefunctions are identical. The factor
$Z_\alpha^2$ is the probability of finding the channel $\alpha$ in the molecular state.
For a weakly bound state these probabilities are 76\% ($D^0\bar D^{0*}$), 10\% ($D^+\bar D^{-*}$),
11\% ($\omega J/\psi$), and 0.8\% ($\rho J/\psi$)\cite{ess} (plus 2\% in D-waves).  Of course 
relative phases of
these components must be accounted for in Eq. \ref{vmd}. The photon momentum is denoted $q$. Finally, the factor of 3 appearing in this expression is a reflection of the famous VMD ratio $\rho:\omega:\phi = 9:1:2$.

\begin{figure}[h]
\includegraphics[angle=0,width=5cm]{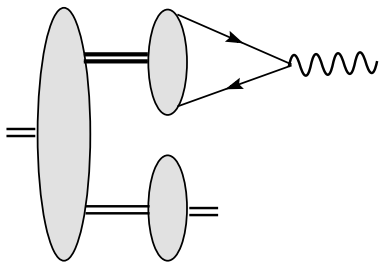}
\caption{\label{VMD} Vector Meson Dominance Diagram}
\end{figure}

\begin{figure}[h]
\includegraphics[angle=0,width=5cm]{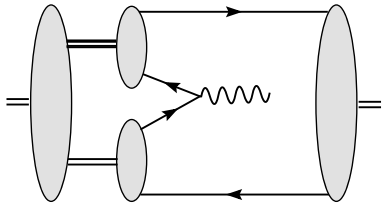}
\caption{\label{ANN} Annihilation Diagram}
\end{figure}

The annihilation contribution to radiative decay illustrated in Fig. 2 proceeds via
the $D^0\bar D^{0*}$ + c.c. and $D^-\bar D^{+*}$ + c.c. components of the $\hat\chi_{c1}$. In 
evaluating this diagram I have found  it convenient to assume simple harmonic
wavefunctions (SHO) for the $J/\psi$, $D$, and $D^*$ mesons. These wavefunctions are
specified in terms of an SHO width parameter, $\beta$ (I also set $\beta_D = \beta_{D^*}$). 
This assumption permits analytic integration over eight of the nine required dimensions.
The remaining integral over the radial $\hat\chi$ wavefunction is performed numerically. 
The final result for the width due to annihilation is 

\begin{equation}
\Gamma_{\rm ANN} = {4\over 27} \alpha {q E_\psi \over m_\chi} {\rm e}^{-q^2\over 2 \beta_D^2} \left( \eta_{00} - {1\over 2} \eta_{+-} \right)^2,
\label{ann}
\end{equation}
where the factors $\eta_\alpha$ are a convolution of the radial $\hat\chi$ wavefunction
in the $\alpha$ channel with the intermediate $D$ and $D^*$ wavefunctions and the
final state charmonium wavefunction. Computations with SHO wavefunctions reveal that
these factors are zero unless the final charmonium state is an S-wave. While this is not
true for more realistic wavefunctions (such as Coulomb+linear) I choose not to evaluate
the integrals in this case since the results will be small and will depend on model
details. Relevant S-wave $\eta$ factors are given by the following expressions:

\begin{equation}
\eta_\alpha(1S) = {\sqrt{8}\over \pi^{3/4}} \left({\beta_\psi \over 2\beta_\psi^2+\beta_D^2}\right)^{3\over 2} \int d^3p \, Z_{\alpha} \phi_\alpha(p)\, \exp\left(-{\rho^2 p^2\over 2\beta_\psi^2+\beta_D^2}\right)
\end{equation}
and

\begin{equation}
\eta_\alpha(2S) = {4\over \sqrt{3}\pi^{3/4}} \left({\beta_\psi \over 2\beta_\psi^2+\beta_D^2}\right)^{3\over 2} \int d^3p \, Z_{\alpha}\phi_\alpha(p)\,  \exp\left(-{\rho^2 p^2\over 2\beta_\psi^2+\beta_D^2}\right)\cdot
\left( {3 \beta_\psi^2\over 2\beta_\psi^2+\beta_D^2} - {4 \rho^2 p^2 \over \beta_\psi^2}
{(\beta_\psi^2+\beta_D^2)^2\over (2\beta_\psi^2+\beta_D^2)^2}\right).
\end{equation}
These expressions depend on the quark mass ratio $\rho = m_c/(m_c+m_u)$.

The annihilation and vector meson dominance amplitudes are evaluated with $\hat \chi$
wavefunctions computed in Ref. \cite{ess} and typical quark model parameters
$\beta_\psi = 0.67$ GeV, $\beta_D = \beta_{D^*} = 0.4$ GeV, $m_c = 1.6$ GeV, and $m_u = 0.33$ GeV. The vector wavefunction at the origin is taken to be $\psi_\omega(0) = 0.976$ GeV$^{3/2}$ as determined by numerically integrating the Schr\"odinger equation with a Coulomb+linear+smeared hyperfine potential. This compares favourably with the SHO result of 1.17 GeV$^{3/2}$. The final rate for $X \to \gamma J/\psi$ is dominated by the VMD and cross terms in
the width and is presented in Table II.  The rate for $X \to \gamma \psi''$ and $\gamma \psi_2$ are zero for SHO wavefunctions since they only proceed via the annihilation diagram. Finally $X \to \gamma \psi'$ is
also driven by the annihilation diagram and is very small.

A variety of predictions for radiative decays of the $\chi'_{c1}$ (assuming a mass of 3872 MeV) are also presented in 
Table II. The fourth column summarises the results of Barnes and Godfrey\cite{BG}.
These rates are computed in the impulse, nonrelativistic, zero recoil, and
dipole approximations. The results for $X \to \gamma J/\psi$ are particularly sensitive
to model details\footnote{This rate is zero for SHO wavefunctions.}. This is examined in
columns five and six which present the results of two additional computations. The first,
labelled [A], employs the same approximations of Barnes and Godfrey but uses meson
wavefunctions computed with a simple, but accurate, Coulomb+linear+smeared hyperfine
potential. It is apparent that the $\gamma J/\psi$ rate is very sensitive to wavefunction
details. Furthermore, one may legitimately question the use of the zero recoil and dipole approximations
for the $\gamma J/\psi$ mode since the photon momentum is so large in this case. The
sixth column (model [B]) dispenses with these approximations, and one finds a relatively
large effect for $\gamma J/\psi$.

\begin{table}
\caption{E1 Decays of the X(3872).}
\begin{tabular}{ccc|ccc|c}
mode & $m_f$ (MeV) & $q$ (MeV)  & $\Gamma[c\bar c]$ (keV) & $\Gamma[c\bar c]$ (keV) & $\Gamma[c\bar c]$ (keV) & $\Gamma[\hat \chi_{c1}]$ (keV)  \\
   &   &   & [B\&G]   &   [A]   &   [B]   &  \\
\hline
$\gamma J/\psi$ & 3097 & 697 & 11 & 71 & 139 & 8 \\
$\gamma \psi'(2^3S_1)$ & 3686 & 182 & 64 & 95 & 94  & 0.03 \\
$\gamma \psi^{''}(1^3D_1)$ & 3770 & 101 &  3.7 & 6.5 & 6.4 &   0 \\
$\gamma \psi_2(1^3D_2)$ & 3838 & 34 & 0.5 & 0.7 & 0.7 & 0 \\
\hline
\end{tabular}
\end{table}

Table II makes it clear that computations of the $\gamma J/\psi$ radiative transition of
the $\chi'_{c1}$ are very sensitive to model details\footnote{There is an additional error induced by arbitrarily changing the quark model $\chi'$ mass to 3872 MeV. This is made clear through the observation
that the dipole formula for the width scales as $q^3$ whereas the momentum space formula scales as $q$.}
. The result of Barnes and Godfrey is
similar to that computed here for the molecular $\hat\chi$ state but is much smaller than
models A and B. Furthermore, the rates for $\gamma \psi^{''}$ and $\gamma \psi_2$ are
very small for a molecular $X$ (at the order of eV) and quite small for a charmonium $X$.
Perhaps the most robust diagnostic is the $\gamma \psi'$ decay mode. For a molecular
$\hat \chi$ this can only proceed via the annihilation diagram of Fig. 2 and hence is
very small. Clearly a measurement of the $\gamma J/\psi$ and $\gamma \psi'$ decay modes
of the $X(3872)$ will provide compelling clues to its internal structure.

The figures in the last column  of Table I correspond to a $\hat\chi$ state with a binding energy of 1 MeV.
Changing the short range cutoff in the pion-exchange interaction of Ref.\cite{ess} 
modifies the binding energy and, hence, the characteristics of the $\hat \chi$. This affects
decay rates, for example a binding energy of 4 MeV increases the $\gamma J/\psi$ rate by roughly
20\% and has minimal impact on the $\gamma\psi'$ rate.

\section{Wavefunction Suppressed Decays: $X \to K\bar K^*$ and $\pi\rho$}

Hidden flavour changing transitions of the $X$ are interesting because they 
occur via $c\bar c$ annihilation at the origin for molecular states and via
multigluon intermediate states for charmonia. One may hope that these very different
mechanisms will provide additional clues to the nature of the $X$.

The lowest order diagram in the strong
coupling in the molecular case is illustrated in Fig. 3. Here the dashed line 
refers to instantaneous gluon exchange. The operator giving rise to this diagram
is taken to be

\begin{equation}
V_{oge} = {1\over 2} \int d^3x d^3y \, \psi^\dagger(\x) T^a \psi(\x) K(\x-\y) \psi^\dagger(\y) T^a \psi(\y)
\label{voge}
\end{equation}
where $T^a$ is a colour generator and $K$ is a kernel given by 

\begin{equation}
K(r) = -{\alpha_s\over r} + {3  \over 4} b r.
\end{equation}
The constants appearing here are the strong coupling and the string tension respectively.
This operator comprises a relativistic extension of the quark model and hence the computations presented here are on 
somewhat less firm ground than the radiative transitions of the previous section. However,
note that the form used here is consistent with the Hamiltonian of QCD in Coulomb
gauge.

\begin{figure}[h]
\includegraphics[angle=0,width=5cm]{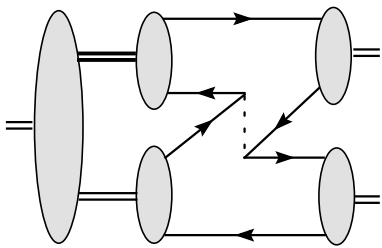}
\caption{\label{SUP} Wavefunction Suppressed Annihilation Diagram}
\end{figure}

It is likely that this diagram is much larger than
the analogous transverse gluon exchange diagram because the latter contains an 
intermediate $u\bar u g$
or $d\bar d g$ hybrid (bound state time ordered perturbation theory is employed throughout
this paper) and hence is suppressed by a large  energy denominator.

The initial state of Fig. 3 is the charged or neutral $D\bar D^*$ component of the $\hat\chi$; this permits the annihilation of the charmed quarks and the production of an $s\bar s$
pair, which materialize in the $K\bar K^*$ final state. Notice that the production of 
charged kaons can only proceed via the neutral, and dominant, $D\bar D^*$ component of the
$\hat \chi$, while neutral kaons arise from the charged $D\bar D^*$ component. Thus one 
expects $\Gamma(X \to K^+\bar K^{-*}) \gg \Gamma(X \to K^0\bar K^{0*})$ for molecular $X$
while $\Gamma(X \to K^+\bar K^{-*}) \approx \Gamma(X \to K^0\bar K^{0*})$ for charmonium $X$.

The computation of Fig. 3 assumes SHO $K$, $K^*$, $D$ and $D^*$ meson wavefunctions as above. The result is

\begin{equation}
\Gamma(\hat \chi_{c1} \to K^-K^{+*}) = {1\over 8}q {E_K E_{K^*}\over m_\chi} \left|{C_F\over 2N} (6b + 8\alpha_s{\beta_K^2\beta_D^2\over \beta_K^2+\beta_D^2}) {1\over m_c m_s} {\beta_K\beta_D\over \beta_K^2+\beta_D^2} \eta_{00}(q) \right|^2,
\label{KK}
\end{equation}
where $C_F = (N^2-1)/(2N)$, $N$ is the number of colours, and $\eta_\alpha$ is a factor defined by

\begin{equation}
\eta_\alpha(q) = \int_0^\infty dp \, {p\over q}\, \phi_{\alpha R}(p) \, \left[ \exp\left(-{(p-q)^2\over \beta_D^2+\beta_K^2}\right) - 
\exp\left(-{(p+q)^2\over \beta_D^2+\beta_K^2}\right) \right].
\end{equation}
The radial wavefunction appearing in this equation is defined by $\phi_\alpha = Y_{\ell m} \phi_{\alpha R}$.

Evaluating Eq. \ref{KK} with typical quark model parameters $b = 0.18$ GeV$^2$, $\alpha_s = 0.5$, and $\beta_K = 0.4$ GeV yields 

\begin{equation}
\Gamma(\hat \chi_{c1} \to K^+K^{-*}) = 4.4 \ {\rm  eV}. 
\end{equation}
The rate to neutral kaons proceeds through the smaller charged $D\bar D^*$ component
of the $\hat \chi$ and is given by

\begin{equation}
\Gamma(\hat \chi_{c1} \to \bar K^0 K^{0*}) = 0.8 \ {\rm eV}. 
\end{equation}

Comparable rates for P-wave charmonia are only known for the $\chi_{c0}$ and are

\begin{eqnarray}
\Gamma(\chi_{c0} \to K^+K^-) &=& 95(28) \ {\rm keV} \nonumber \\
\Gamma(\chi_{c0} \to K^0_sK^0_s) &=& 32(14) \ {\rm keV}.
\end{eqnarray}
Using these figures as a benchmark for $2^3P_0 c\bar c \to K\bar K^*$ indicates that
multigluon annihilation of charmonium $\chi'_{c1}$ is roughly 1000 times larger
than the wavefunction suppressed decay of the molecular $\hat\chi$.  Thus finding the
$X$ in $K\bar K^*$ at the keV level may be an indication that it is a charmonium and not a molecular state.

The same computation applies to the $\pi\rho$ decay mode with the exception that the strange
quark mass is replaced with the up or down constituent mass, the kinematics are slightly
different, and both the charged and neutral $D\bar D^*$ components of the $\hat\chi$ wavefunction
contribute to $\pi^+\rho^- + {\rm c.c.}$ and $\pi^0\rho^0$ decays. Taking $\beta_\pi$ = $\beta_\rho = 0.4$
GeV yields

\begin{equation}
\Gamma(\hat \chi_{c1} \to \pi^+\rho^- + {\rm c.c.}) = \Gamma(\hat \chi_{c1} \to \pi^0\rho^0) = 40\  {\rm eV}.
\end{equation}
One expects that the analogous charmonium rates are several orders of magnitude greater than this
so that discovering the $X$ in the $\pi\rho$ channel could be an indication that it is a
charmonium state.

\section{Conclusions}

Radiative decays of the $X(3872)$ offer a promising method to distinguish the 
charmonium and molecular assignments for this state. 
Unfortunately the predicted rate for charmonium $2^3P_1$ to $\gamma J/\psi$ is 
very sensitive to model details so that a comparison to the molecular rate
is less significant that desired. Nevertheless, one expects $\Gamma(\hat\chi_{c1}\to \gamma J/\psi) 
\alt \Gamma(\chi'_{c1} \to \gamma J/\psi)$. Furthermore, determining this width for the genuine
charmonium state will provide a demanding test of our understanding of heavy quark systems.

Alternatively, the charmonium width for $\gamma \psi'$ is relatively stable and is much
larger than the molecular rate, which can only proceed via light quark annihilation, and is
therefore wavefunction  suppressed. Thus collecting sufficient $X$ events to verify that the $\gamma \psi'$ final
state is not seen at  the 100 keV level will provide compelling evidence of the molecular
nature of the $X$.

It has been argued that existence of $2\pi J/\psi$ and $3\pi J/\psi$ decay modes of the $X$
are a strong indication of the isospin symmetry violating, and hence molecular, nature of the $X$.
In a similar fashion, the suppressed hidden flavour changing decay $X \to K\bar K^*$ strongly favours
the  charged kaon final state over neutral kaons. Alternatively, one expects equal production
of charged and neutral kaons from charmonium decay. An explicit computation shows that the molecular
rate is very small, and probably several orders of magnitude smaller than the analogous charmonium
decay. Thus a direct measurement of this rate or of the ratio of charged to neutral kaons may be
a useful test of the nature of the $X(3872)$.

\begin{acknowledgments}
I thank Ted Barnes, Eric Braaten, Olga Lakhina, Philip Page, Walter Toki, and the participants of
the BaBar collaboration workshop on charmonium  for discussions.
This work was supported by the DOE under contract DE-FG02-00ER41135.
\end{acknowledgments}

\end{document}